\documentclass[a4paper, 11pt, twocolumn]{article}
\usepackage[utf8]{inputenc}
\usepackage{lrec}

\usepackage{csquotes}
\usepackage{multicol}

\usepackage[T1]{fontenc}
\usepackage{Alegreya}
\usepackage{flushend}
\usepackage{titlefoot}

\usepackage{float}
\newfloat{footnote}{hb}

\usepackage{hyperref}

\usepackage{fancyhdr}

\title{\textbf{Ground Truths for the Humanities $^*$}}
\author{Yvette Oortwijn$^\dagger$ $^\ddagger$, Hein van den Berg $^\dagger$, Arianna Betti $^\dagger$}
\date{\small{October 25, 2020} \\

\vspace{1em}
\rule{\textwidth}{1pt}
\large{\textbf{Abstract}}
\vspace{-0.5em}
\begin{flushleft}
\footnotesize{Ensuring a faithful interaction with data and its representation for humanities \textit{can} and \textit{should} depend on expert-constructed ground truths. \\}
\rule{\textwidth}{1pt}
\end{flushleft}
\vspace{-2em}}

\fancypagestyle{firststyle}
{
\fancyhf{}
\fancyfoot[L]{\begin{tabular}[b]{@{}p{\linewidth}@{}}
    \vspace{0.8em}
    \rule{\textwidth}{1pt}
    \scriptsize
    $^*$ Provocation paper presented at the 5th Workshop on Visualization for the Digital Humanities (VIS4DH), part of the 31st IEEE Visualization Conference, IEEE VIS 2020, Virtual Event, Salt Lake City, USA, October 25-30, 2020 \\ \vspace{-1em}
    \scriptsize $^\dagger$ University of Amsterdam, Institute for Logic, Language and Computation $^\ddagger$ Eindhoven University of Technology, Algorithms, Geometry \& Applications\\\vspace{-1em}
    \scriptsize \begin{flushright}This work is licensed under the arxiv.org license. License details: \url{https://arxiv.org/licenses/nonexclusive-distrib/1.0/license.html} \end{flushright}\\ \footnotesize\vspace{5pt}
\centering \thepage
\end{tabular}
}}

\begin{document}


\maketitle

\thispagestyle{firststyle}

\section{Provocation statement \& argument}
An important measure for the faithfulness of representations of data is the comparison to a ground truth. 

According to some, ground truths are unattainable in humanities research (\newcite[1]{kirschenbaum_remaking_2007}; \newcite[8-9]{Nguyen_2020}). But the fact that humanities research is concerned with “interpretation, ambiguity and argumentation” \cite[1]{kirschenbaum_remaking_2007} does not imply, we submit, that ground truths for humanities data are impossible or inessential.

Expert knowledge-based evaluation of computational representations is necessary because generic benchmarks fail to guarantee reliable results for downstream tasks on specialised data (\newcite{gladkova-drozd-2016-intrinsic}; \newcite{bakarov2018survey}). And we say that wherever expert knowledge is available, ground truths can be constructed \cite{vandenberg2018philosophical}, in any field of knowledge.

We have developed a method for constructing ground truths in any concept-focused textual domain \cite{betti:20}. The method relies on our \enquote{model approach} for fixing the interpretation of a concept, where concepts are represented by complex, networked relations between terms \cite{betti2014modelling}. Our models can be easily turned into schemes for annotating textual fragments. Annotations can be used to test whether the output of computational analysis matches the best-supported interpretation of fragments, increasing the objectivity and replicability of humanities research. Domains in humanities which do not focus on concepts should develop similar methods for constructing ground truths.

\section{Counter-perspective}
We read \newcite{Nguyen_2020} as suggesting that measures of reliability alternative to ground truths are necessary because multiple valid definitions of concepts exist in the humanities \cite[8-9; 17-18]{Nguyen_2020}. It is naive, though, to assume the sciences to be different: definitions of concepts are always interpretations (\cite{Laplane3948}; \cite{vandenberg2018philosophical}). Interpretations in the humanities are bound by criteria of objectivity: they must be textually adequate, and preferable to other interpretations, e.g. when they explain a text better than other ones. Computational textual representations can be evaluated on the basis of the best models available: a procedure common to many sciences.

\section*{\centering Acknowledgements}
This research was supported by grants e-Ideas (VICI, 277-20-007) and CatVis (314-99-117), funded by the Dutch Research Council (NWO), and by the Human(e)AI grant Small data, big challenges funded by the University of Amsterdam.

\begin{table*}[t]
  \centering
  \begin{tabular}{lcr}
    \large{\textbf{References}}    
  \end{tabular}
\end{table*}
\bibliographystyle{lrec}
\bibliography{references}
 \par\leavevmode
 \end{document}